\documentclass{iau}
\usepackage{graphicx,natbib}
 
\newcommand{\mnras}{MNRAS}       

\newcommand{\aap}{A\&A}

\newcommand{\aj}{AJ}

\newcommand{\apjs}{ApJS}           

\title{3D spectroscopy of Wolf-Rayet HII galaxies}
\author[Kehrig et al.]{C. Kehrig$^1$, E. P\'erez-Montero$^{1}$,
  J.M. V\'ilchez$^{1}$, J. Brinchmann$^{2}$, D. Kunth$^{3}$,
  F.Durret$^3$, J.Iglesias-P\'aramo$^1$ \and  J.Hern\'andez-Fern\'andez$^4$}

\affiliation{$^1$Instituto de Astrof\'isica de Andaluc\'ia  - Apartado de correos 3004, 18080 Granada, Spain \\
$^2$Leiden University - PO Box 9513, 2300 RA Leiden, The Netherlands\\ 
$^3$Institut d'Astrophysique de Paris - 98 bis boulevard Arago, 75014 Paris, France 
$^4$Universidade de S\~ao Paulo (IAG) - Rua do Mat\~ao 1226,
05508-090 SP, Brazil}
\pubyear{2014}
\volume{309}
\jname{Galaxies in 3D across the Universe}
\editors{B. L. Ziegler, F. Combes, H. Dannerbauer, M. Verdugo, eds.}

\begin{document}

\maketitle

\begin{abstract}
  Wolf-Rayet HII galaxies are local metal-poor star-forming galaxies,
  observed when the most massive stars are
  evolving from O stars to WR stars, making them template systems to
  study distant starbursts. We have been performing a program to
  investigate the interplay between massive stars and gas in WR HII
  galaxies using IFS. Here, we highlight some results from the
  first 3D spectroscopic study of Mrk~178,  {\it the closest metal-poor WR HII
   galaxy}, focusing on the origin of the 
  nebular HeII emission and the aperture effects on the detection of WR features.


\end{abstract}

\firstsection
\section{Introduction}

HII galaxies are local, dwarf starburst systems
\citep[e.g.,][]{K04,W04}, which show low metallicity [1/50 $\lesssim$
Z/Z$_{\odot}$ $\lesssim$ 1/3]\citep[e.g.,][]{K06,EPM09,C09,C10}. Wolf-Rayet
(WR) signatures (commonly a broad feature at $\sim$ 4680 \AA~ or
blue bump), indicating the presence of WR stars, have been found in
the spectra of some HII galaxies \citep[e.g.,][]{KS81,K13}. This is an
important observational fact since according to recent stellar
evolution models for single rotating/non-rotating massive stars,
hardly any WRs are expected in metal-poor environments
\citep[][]{L14}. Studying the WR content in HII galaxies is crucial
to test stellar evolutionary models at low metallicities. We have
initiated a program to investigate HII galaxies with WR features using
integral field spectroscopy (IFS; e.g., \citealt{K13,EPM13}). So far,
we have observed 15 WR galaxies with the optical IFUs: PMAS at the
3.5m telescope at CAHA and INTEGRAL at the 4.2m WHT in ORM. IFS has
many benefits in a study of this kind, in comparison with long-slit
spectroscopy. Using IFS one can locate and find WRs where they
were not detected before, not only because it samples a larger area of
the galaxy, but also because IFS can increase the contrast of the WR
bump emission against the galaxy continuum, thus minimizing the WR
bump dilution. Also, IFS is a powerful
technique to probe issues related with aperture effects, and allows a
more precise spatial correlation between massive stars and nebular
properties \citep[e.g.,][]{K08,K13}.

\section{Results and Conclusions}

\begin{figure}
\centering
\includegraphics[width=0.54\textwidth]{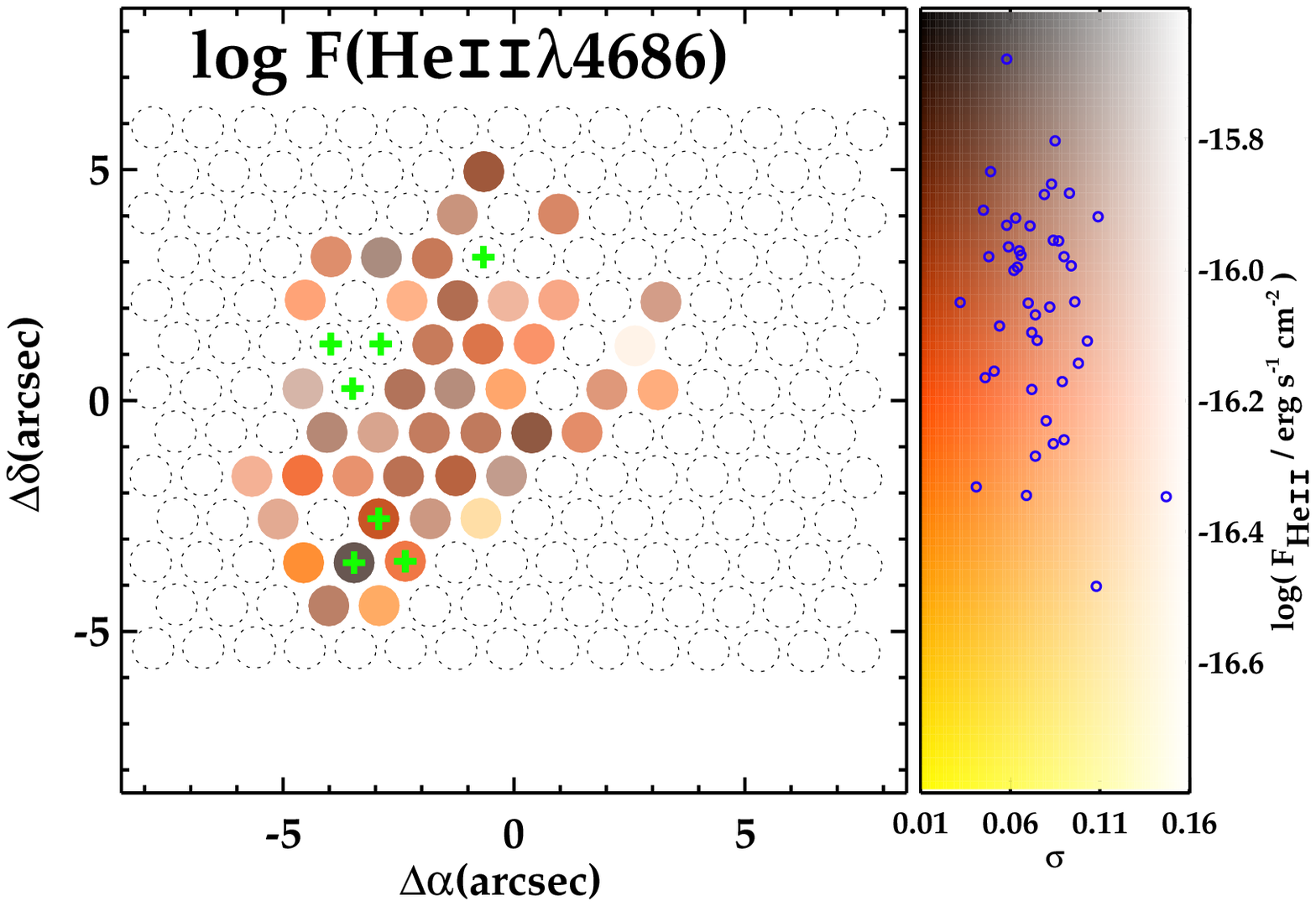}
\includegraphics[width=0.45\textwidth]{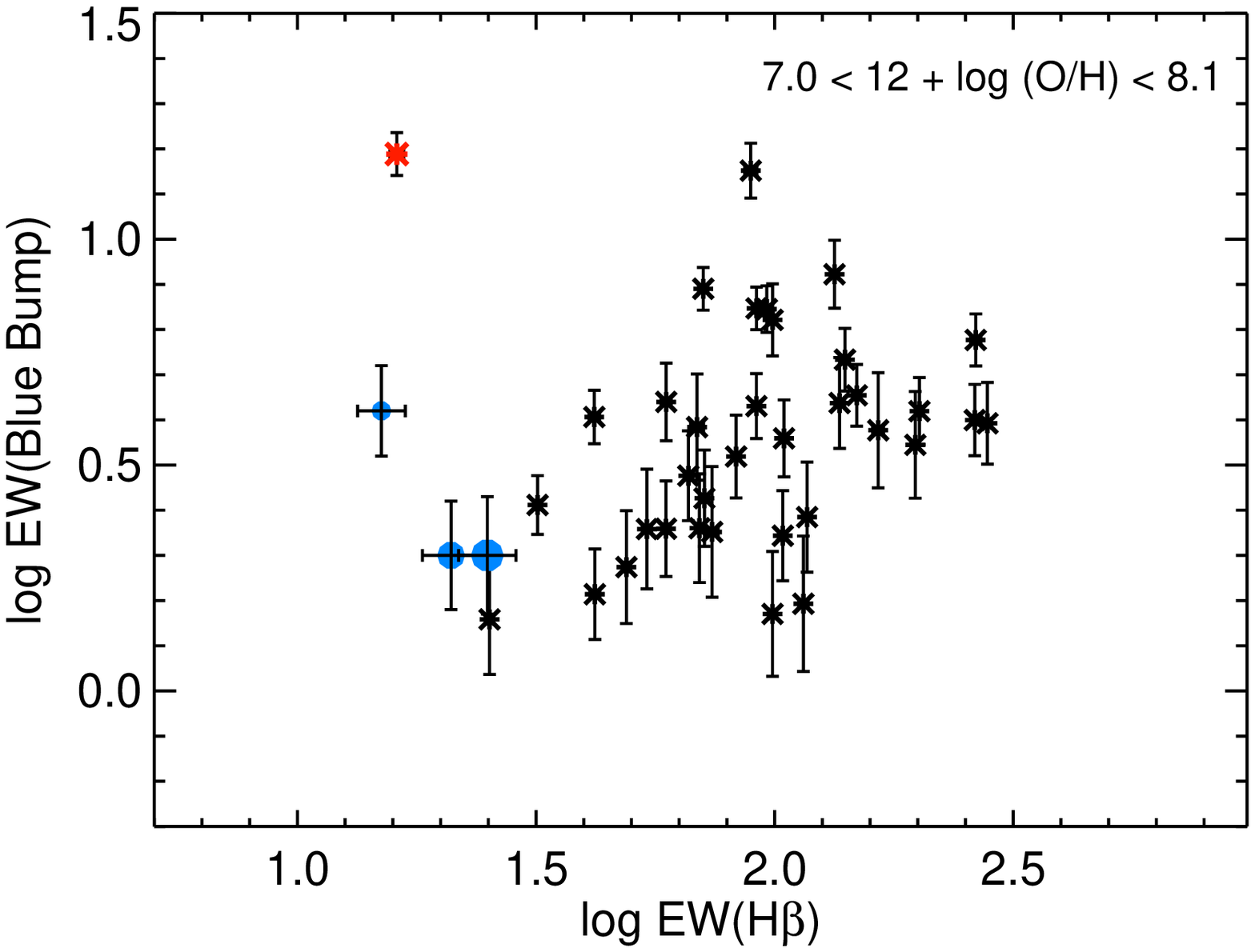}
\caption{Mrk178 - {\it Left panel}: intensity map of the nebular HeII$\lambda$4686
  line; the spaxels where we detect WR features are marked with green
  crosses. {\it Right panel}: EW(WR blue bump) vs EW(H$\beta$). Asterisks show values
  obtained from SDSS DR7 for metal-poor WR galaxies; the red one represents
  Mrk178. The three blue circles, from the smallest to the biggest
  one, represent the 5, 7 and 10 arcsec-diameter  apertures from
our IFU data centered at the WR knot of Mrk178, at which the SDSS fiber
was centered too~\citep{K13}.}
\label{fig}
\end{figure}

We summarize here some recent results on Mrk178, one of {\it the
most metal-poor nearby WR galaxies} \citep[see][for more details]{K13}:

1) The origin of high-ionization nebular lines (e.g. HeII$\lambda$4686), apparently more frequent in high-z galaxies, is still an
open question. One widely favored mechanism for He$^{+}$-ionization involves hot WRs, but it has been shown that nebular
HeII$\lambda$4686 is not always accompanied by WR signatures, thus
WRs do not explain He$^{+}$-ionization at all times
\citep{K08,K11,SB12}. In Mrk178, we find nebular HeII$\lambda$4686
emission spatially extended reaching well beyond the location of the
WR stars (Fig.~\ref{fig}, left-panel). The excitation source of He$^{+}$ in Mrk178 is
still unknown.

2) From the SDSS spectra, we have found a too high EW(WR bump) value for
Mrk178, which is the most deviant point among the metal-poor WR
galaxies in Fig.~\ref{fig}, right-panel. Using our IFU data, we
have demonstrated that this curious behaviour is caused by aperture effects,
which actually affect, to some degree, the EW(WR bump) measurements for all galaxies in
Fig.1. We have also shown that using too large an aperture,
the chance of detecting WR features decreases, and that WR signatures
can escape detection depending on the distance of the object and on
the aperture size. Therefore, WR galaxy samples/catalogues constructed
on single fiber/long-slit spectrum basis may be biased!

\section*{Acknowledgements}

This work has been partially funded by research project AYA2010-21887-C04-01 from the Spanish PNAYA.

\end{document}